\documentstyle[]{mn}
\input psfig.sty

\begin{document}
 
\title{Evidence for evolving accretion disk structure in 4U1957+115}
\author[P.J. Hakala, P. Muhli, G. Dubus]
       {P.~J. Hakala$^{1,2}$, P. Muhli$^{1}$, G. Dubus$^{3,4}$ \\
        $^1$ Observatory and Astrophysics Laboratory, P.O. Box 14, FIN-00014 
          University of Helsinki, Finland\\
        $^2$ Mullard Space Science Laboratory, University College
        London, Holmbury St Mary, Dorking, UK\\
 	$^3$ DARC, UPR 176 du CNRS, Observatoire de Paris Meudon, 
        5 place Janssen, 92195 Meudon, France\\ 
        $^4$ Astronomical Institute "Anton Pannekoek", Kruislaan 403, 1098 SJ 
        Amsterdam, the Netherlands\\ 
}
\date{November 1998}
 
\maketitle
\begin{abstract}

We present results of the first multicolour photometric study of 4U1957+115,
one of the optically less studied low mass X-ray binaries (LMXB).
Our quasi-simultaneous UBVRI observations 
reveal that the light curve pulse shape over the 9.33 hr period discovered by \cite{thorst}
has changed substantially since his earlier V band studies. The light curve now shows
clear asymmetry and an amplitude more than double of that seen by \cite{thorst}. 
The light curve also shows colour dependence that seems to rule out the X-ray heating 
model as an origin for photometric variability. We believe that the
changes observed in the light curve shape are indicative of evolving
accretion disk structure.  We discuss implications of this on the 
origin of optical emission in 4U1957+115 and disk accreting binaries in general. 

\end{abstract}

\pagebreak

\section{Introduction}

Low mass X-ray binaries (LMXB's) are semi-detached compact interacting 
binaries, where a late type secondary star loses matter via Roche lobe
overflow. This matter forms an accretion disk around a compact primary, which
is either a neutron star or a black hole. The optical emission from
LMXB's originates in both the accretion disk itself and the X-ray 
heated secondary star. The accretion disk typically
dominates the emission all the way from the optical through UV to the
X-rays. Unlike in cataclysmic variables, the optical emission from
accretion disks in LMXB's is mainly due to reprocessing of the X-rays
in the disk and not viscous heating. In many cases the optical light 
curve is roughly sinusoidal. This can be attributed to the 
X-ray heated secondary. In these
systems, the source is generally bluest during the flux maximum. This
is due to the fact that the X-ray heated face of the secondary is much
hotter than the other side of the companion star. This is the case in
systems like X1254-69, Sco X-1, GX339-4, GX9+9 and X1735-444
\cite{vanpara}. In some other systems, like accretion
disk corona (ADC) sources X1822-371 and AC211, the origin of
the optical light curves has generally been explained differently
\cite{hellmason}, \cite{callanan}. The favoured view has been that the 
optical emission from the accretion disk is the main source of optical 
radiation, and that it is modulated in relatively high inclination 
systems by the vertically extended outer rim of the disk 
(see for instance \cite{mascor82}, \cite{hellmason}).   

4U1957+115, as the name suggests, was first detected by the {\it Uhuru}
mission \cite{giacc74} and its optical counterpart was later identified by
\cite{margo78}. The source also appears in the early list of black hole 
candidates \cite{whma85}, as it shows an ultrasoft X-ray
spectrum. However, as of now, there is no firm evidence on the nature
of the compact object. The source has been noted to vary in the
optical \cite{motch85}, but there has not been any further optical photometry
published, since \cite{thorst} discovered a 9.33 hrs modulation in his 
V band observations. This period, if of orbital origin, implies that
if the secondary fills its Roche lobe and belongs to the main
sequence, it should have a mass of 0.11*P$_{orb}$ = 1.0 M$_{\odot}$.
There is, however, no firm sign of the secondary in the optical
spectra \cite{shahbaz}.

In this paper we report first multicolour observations of 4U1957+115
and present an approach for fitting the optical light curve and thus
extract some information on the 3D shape of the disk.  Section
2 gives details of our observations as well as data reductions performed. 
In section 3 we give a description of our model and present the
results of the model fitting. Finally, in section 4, we discuss the
implications of our modelling both in relation to 4U1957+115 and
LMXB's in general.       

\section{Observations}

4U1957+115 was observed on Nordic Optical Telescope (NOT), Observatorio de 
Roque de los Muchachos, La Palma on 9-10th of August, 1996. NOT is a 2.56m
Cassegrain telescope and at the time of our observations, it was fitted with
a 2kx2k thinned Loral-Lesser CCD detector. The observations were carried out as a 
continuous sequence automatically cycling U,B,V,R,I filters during
the two consecutive nights. Our data only cover one 9.33 hour period,
and the resulting folded light curves thus have no phase overlap.

The data were debiased and flatfielded in usual manner, after which digital
aperture photometry was performed. this was done using the DAOPHOT
routines available for IDL. The delta magnitudes were measured against
a comparison star in the same field of view (star 6 in Doxsey et
al. (1977)). These delta magnitudes where then converted to
absolute magnitudes by observations of a relatively blue
standard star number 108 551 from \cite{landolt}.

The resulting U,B,V,R,I light curves are plotted in
Figure 1. As the data is not simultaneous in each filter, but we
cycled the UBVRI filters throughout the observations, we
used 4th order Fourier fits to characterize the data and to measure
the colour indices (Figure 2.).
As a consistency check we also measured the B magnitude difference between star 6
and another star of similar brightness about 20" NW of star 6. We found no systematic
effects and the scatter in that data was 0.01 mag (1 $\sigma$). This
is in very good agreement with the errors we get for 4U 1957+115. We
should stress though, that as our absolute magnitude calibration was
only done using a single standard star, this error level is valid only
for the differential magnitudes, not for the zero point calibration.

\begin{figure}
\centerline{\psfig{figure=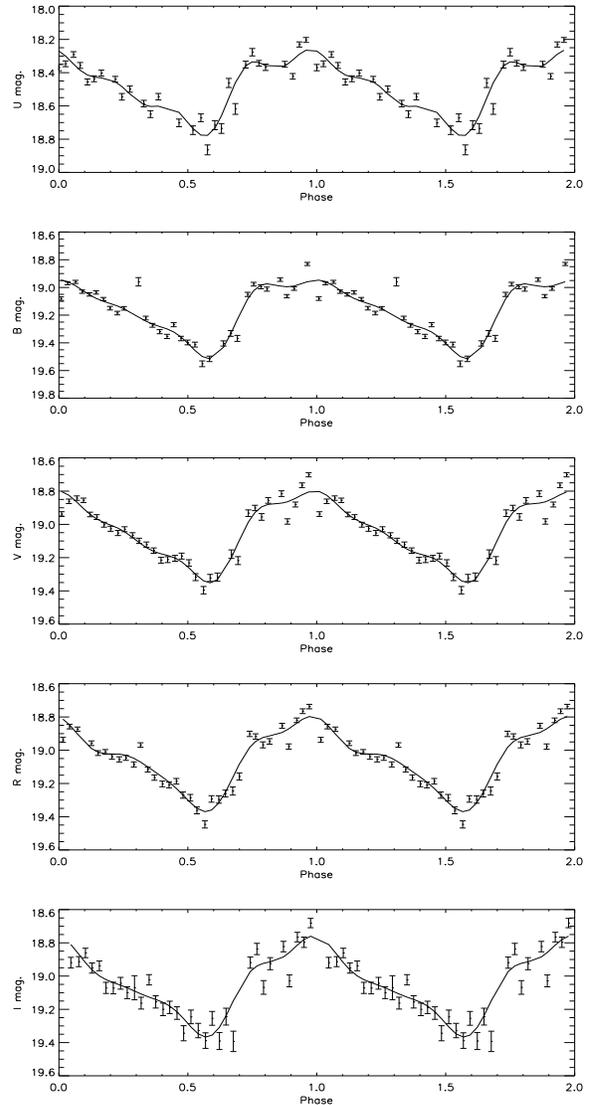,height=15.0cm,width=8.0cm,angle=0}}
\caption{U,B,V,R,I (top to bottom) light curves folded over the 9.33 hrs period. The phase
0.0 is arbitrary. The solid line gives the 4th order Fourier fit used
to measure the variation amplitudes and the colour indices (Figure 2.) 
The data are plotted twice for clarity. The presented errors are valid
for differential photometry only (see text).}
\label{fig:photom}
\end{figure}

\begin{figure}
\centerline{\psfig{figure=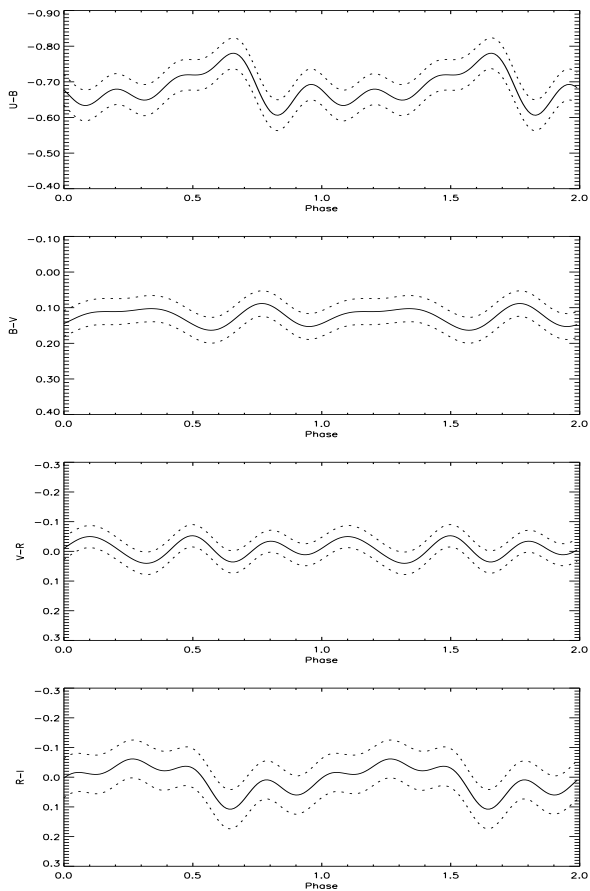,height=12.0cm,width=8.0cm,angle=0}}
\caption{Colour indices as a function of 9.33 h period phase. The phasing is the same
as in Figure 1. The plots show the differences of the fits to the lightcurves shown in
Figure 1. From top to bottom: U-B, B-V, V-R and R-I. In addition to
the fits we have also plotted the 3$\sigma$ upper and lower limits for
the fits (dotted lines).}
\label{fig:colours}
\end{figure}


The only previously published photometric datasets, that we are aware of are
those of \cite{motch85} and \cite{thorst}. \cite{motch85} report a 3.5 hours V band
observation, which suggests variability at timescales of $\sim$ 1 hour with an 
amplitude of $\sim$ 0.1 mag. Later dataset by \cite{thorst} shows very 'clean' sinusoidal 
modulation with the period of 9.33 hours and a peak to peak amplitude of 
0.232 $\pm$ 0.008 mag. This was explained to be due to the X-ray heated 
surface of the secondary, which is otherwise assumed to be of almost solar
type star in terms of mass and radius. Such variability is a common explanation 
for optical light curves in LMXB's. 

\begin{figure}
\centerline{\psfig{figure=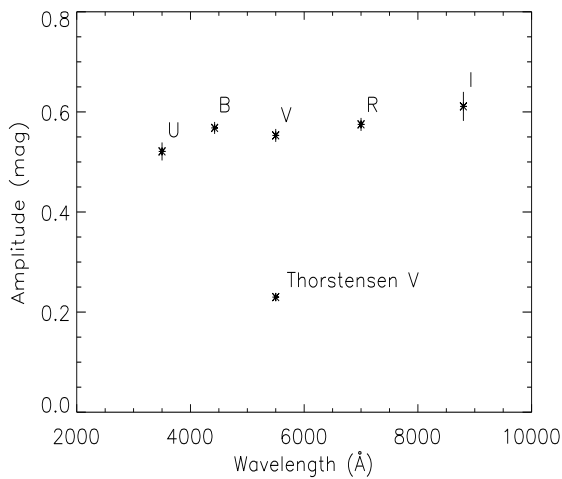,height=7.0cm,width=8.0cm,angle=0}}
\caption{Modulation amplitudes of our dataset as a function of wavelength.
 We also show an earlier V band amplitude measurement by Thorstensen (1987).}
\label{fig:amps}
\end{figure}

Our multicolour data, however, does {\it not} support such hypothesis. It is clear
from our light curves folded on the 9.33 hours period, that the pulse shape is
not sinusoidal. Furthermore, the amplitude of modulation is roughly about twice
that seen by \cite{thorst}. We have estimated variation amplitudes from our  
4th order Fourier fits. These fits are shown, together with our data
in Figure 1. The resulting full UBVRI amplitudes, as a function of
wavelength, are plotted in Figure 3. In addition to these, we have
added a point, which shows the V band amplitude estimate by \cite{thorst}. 

Characterising signatures of X-ray heating model include almost sinusoidal pulse shape
and colour dependence over the orbital period. Typically, the source is bluest
at flux maximum, when the X-ray heated surface of the secondary is facing towards the observer. 
Unfortunately the data published by \cite{thorst} only includes V band. Therefore, they 
could only base the X-ray heating argument on the sinusoidal pulse shape. It is quite
clear, however, from our multicolour light curve (Figure 1. and Figure 2.) that the source is
at its bluest near the phase of {\it minimum} brightness, which totally contradicts the
X-ray heating hypothesis. The colour dependence is only clearly evident in U-B, while
other colours do not show very significant variations over the orbital
phase (Figure 2.). 

In order to produce light curves, which turn bluer when the source
gets fainter, one needs to obscure some of the 'redder' light from the
source. One way to explain this, is to have a grazing eclipse, where
part of the disk outer rim, close to the L1 point gets eclipsed by the
secondary. This has lead us to experiment with disk rim models in
order to explain our results.   

\section{Modelling}

One way to explain light curves of disk-accreting systems
is to assume that the bulk of modulation is due to the varying viewing
aspect of the three dimensional disk. This has especially been applied
to the accretion disk corona (ADC) sources, where \cite{hellmason}
have managed to fit the optical and X-ray light curves of X1822-37
simultaneously by varying the disk rim height over the azimuth. Also,
more recently, \cite{meyhof} have reproduced light curves of supersoft
sources by the same approach. Their models have implied H/R as high as
0.5. Observational evidence for extended vertical structure can also
be found in AC211 and X1916-05 \cite{callanan}, \cite{grindlay}.  

However, there is serious objection to the disk rim models on the
theory side. Even to produce H/R ratios of the order of 0.1,
outer rim of the disk temperatures of the order of 10$^{5}$ K
are required. This is not feasible within our current
understanding of accretion disks in accreting binaries. 
Alternatively, since the discovery of the 35 d period in Her X-1
\cite{tananbaum}  several authors have proposed thin tilted and/or warped
disks to explain observable long term periodicities in X-ray binaries
(see for instance \cite{malo97} for summary of tilted/warped disk models).   

\begin{figure*}
\centerline{\psfig{figure=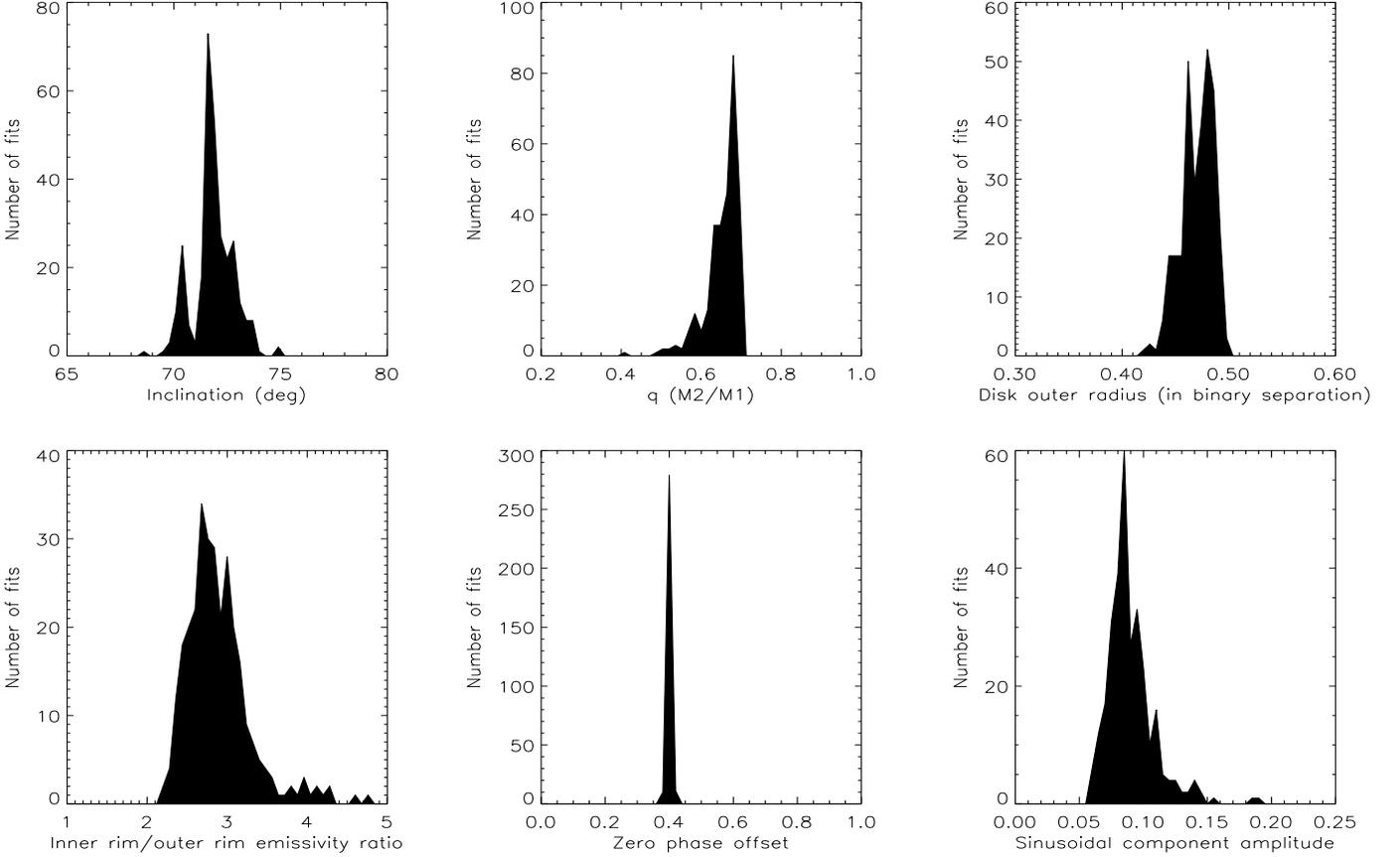,height=12.0cm,width=19.0cm,angle=0}}
\caption{The histograms of system parameter distributions from the
300 independent fits started with different random generations. The
skewed distributions in q and D$_{rad}$ are due to hard upper limits
of 0.7 and 0.5 respectively.}
\label{fig:amps}
\end{figure*}     

In our attempt to model the light curves of 4U1957+115 we have
roughly followed the disk outer rim fitting idea of \cite{hellmason}, 
but with some modifications. Our code proceeds as follows. 

We assume an optically thick disk that has constant surface brightness 
and which is surrounded by vertically extended outer rim. This rim emits both on
its inner and outer surface. The inner surface has the same emissivity
as the disk surface, but the ratio of the rim outer surface emissivity
to that of the inner surface emissivity is a free parameter (as in 
\cite{hellmason}). As there is significantly different colour
information only in the U-B index, we use U and B light curves only.
These are fitted simultaneously. In order to be able to reproduce 
the colour information, we must fix temperatures for the inner and
outer rim. This enables us to compute relative differences in the
emissivity ratios for U and B bands. In our model we use 15000K and
7500K for inner and outer rim respectively. This implies that the U
band emissivity ratio is roughly 1.5 times the B band ratio, for any
fitted value. This values is, however, not very strongly dependent on chosen
temperatures. Following \cite{hellmason} the possible emission
from the X-ray heated face of the secondary is treated as an extra
sinusoidal component in the fit. The amplitude (or relative flux
fraction) of this component is a free parameter in the fit, whilst
the phasing is fixed so that the maximum flux always occurs at the
time when the X-ray heated face of the secondary faces the observer
(phase 0.5). The free system parameters in the fit are:

\begin{itemize}

\item{The ratio of outer/inner rim emissivity, $\nu$}

\item{Inclination of the system, $i$}

\item{Mass ratio $q$, defined as $q=M2/M1$ , where $M1$ is the mass of the
compact object}

\item{Disk outer radius, $D_{rad}$}

\item{Time of eclipse, $\phi_{0}$. (One should note that for 4U1957+115
there is no ephemeris available, thus we do not know the phasing
of the data and this has to be treated as a free parameter as well)}

\item{Amplitude of the sinusoidal (X-ray heating) component, $a$}

\end{itemize}

In addition to these system parameters we do a regularised inversion 
of the disk rim height over the azimuthal direction. To do this, 
we have defined a grid of 36 points spaced by
10 degrees along the azimuthal direction at the outer edge of the
disk. The vertical height of the disk rim at these points is a 
free parameter at each of these locations. However, we do employ a
Tikhonov regularisation functional \cite{tikhonov},\cite{potter}
instead of entropy functional 
in order to find the smoothest possible disk rim that can fit our data.

The actual fitting procedure is carried out by a genetic algorithm
(GA) (see for instance \cite{charbonneau},\cite{hakala},\cite{potter}
for applications of genetic algorithms to astronomical problems).
The merit function to minimise takes the following form.

\begin{equation}
F(\nu,i,q,D_{rad},\phi_{0},a,p_{j}) = \chi^{2}+\lambda\sum_j(p_j-p_{j-1})^2+(p_j-p_{j+1})^2
\end{equation}   

where $\nu$ is the outer rim/inner rim emissivity ratio, $i$ is the
system inclination, $q$ is the mass ratio, $D_{rad}$ is the disk outer
radius, $\phi_{0}$ is the zero phase offset, $a$ is the sinusoidal
component amplitude and p$_{j}$ are the rim heights. $\chi^{2}$ is
the fit Chi-square value, $\lambda$ is the Lagrangian multiplier and the
latter sum is the Tikhonov regularisation term.

Ideally our modelling would come out with best value for all the free 
parameters, but given the number of free parameters and the nonlinear
nature of the optimisation problem it is not immediatedly clear
whether the solutions obtained are unique. Thus, it is vital to test
the uniqueness by perfoming a sample of independent fits. This can be done very easily
with the genetic optimisation method, which intrinsincly always starts 
with a set of random solutions to the problem. Thus, running the 
fitting procedure with different seeds for the random number generator
will produce a set of independent solutions that can be studied for uniqueness.
In Figure 4. we have plotted the histograms of system parameters
obtained from 300 independent fits performed this way. Furthermore, we have
plotted the resulting 300 outer rim profiles in Figure 5.

It is clear from Figure 4. that some of the system parameters are much
better constrained by our modelling than others. Especially, the phase
of the eclipse (i.e. when the compact star is furthest from the
observer) is very well constrained. On the other hand, system
inclination, mass ratio q and the inner rim/outer rim emissivity ratio   
are not that tightly constrained. The same applies to the outer disk
radius as well. More importantly, however, the disk rim shape is
rather well constrained. This is probably not very surprising, since
the bulk of the light curve shape is mostly influenced by the 3D shape
of the disk.

We show a typical fit to the U and B data and in Figure 6. Our model
is capable of producing all the main features in the light curve,
including the difference in the depth of the minimum, which causes
the source to turn bluer in the minimum. In Figure 7. we have plotted
separately the different components of our U band model, shown in 
Figure 6. The light curve is dominated by a roughly sinusoidal
component arising from the variable obscuration of the disk
and inner rim surface by the rim itself. In addition, the outer rim provides the means for
producing the blue minimum, as it shows a wide eclipse by the
secondary. The sinusoidal, X-ray heating, component is found to 
contribute only of the order of 10 \% of the total flux in maximum.     

\begin{figure}
\centerline{\psfig{figure=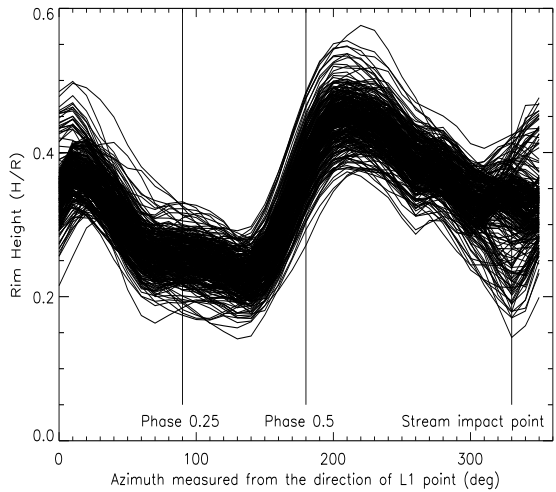,height=7.0cm,width=8.0cm,angle=90}}
\caption{The different rim profiles obtained from our 300 different
fits. The scatter in the plot includes both the effects of the
noise in our data and the possible non-uniqueness in the
solutions. The azimuth is measured starting from the L1 point and
increasing towards the opposite direction to the disk rotation.
The orbital phases at which a particular rim element is between the
compact object and the observer are indicated.}
\label{fig:amps}
\end{figure}     

\begin{figure}
\centerline{\psfig{figure=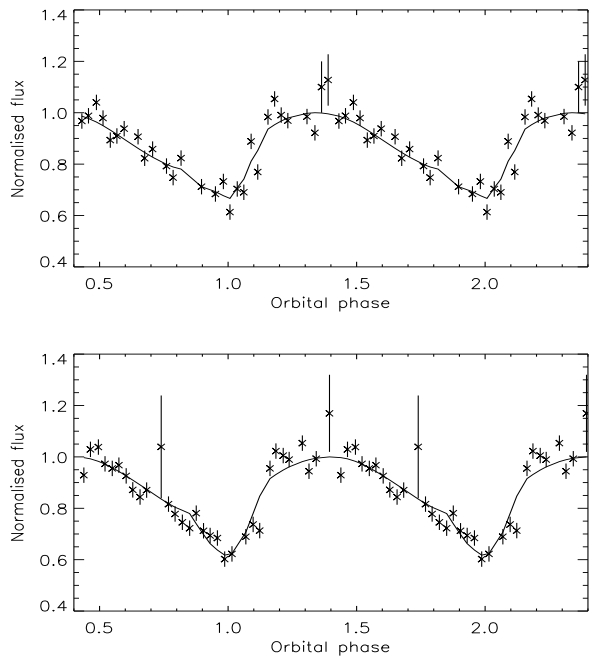,height=12.0cm,width=12.0cm,angle=0}}
\caption{The simultaneous disk model fits to our U band (top) and 
B band light curves. A couple points clearly deviating from the
general trend in the light curve were assigned artificially large
errors to eliminate their contribution to the fit.}
\label{fig:amps}
\end{figure}     

\begin{figure}
\centerline{\psfig{figure=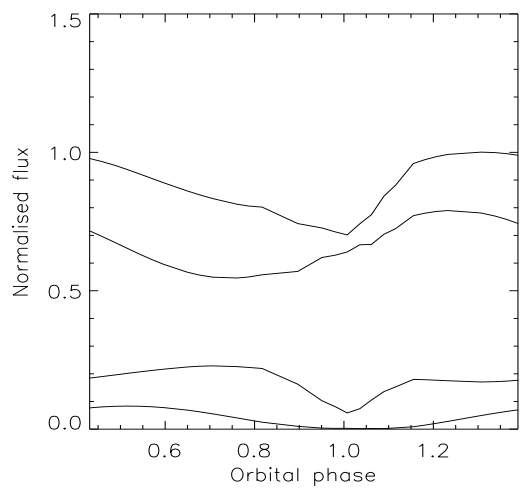,height=7.0cm,width=8.0cm,angle=0}}
\caption{The different emission components that effect the fit. From
top to bottom: The U band fit to the data, contribution from the disk
surface and inner rim, contribution from the outer rim and the
contribution from the X-ray heating component.}
\label{fig:amps}
\end{figure}     

\section{Results}

As a result of our fits, we can set some limits on the
systems parameters of 4U 1957+115. However, we must emphasize that
although our data is of high S/N, these fits are based on a single 
light curve and thus should be noted with caution. 

We find that the system inclination is likely to be 70-75
degrees. This is in agreement with the fact that no clear optical or
X-ray eclipses have been detected. Furthermore, this can explain
the 'blue minimum' as a result of partial eclipse of the accretion
disk. We can also give a lower limit for the mass ratio q, of the
order of 0.4-0.5. Our fits tended towards very high values of q,
and this can be seen as a very skewed distribution in Figure 4.
In our fits we used a hard upper limit of 0.7 for q. This was thought
to be justified as the compact object (as a black hole or neutron
star) has a minimum mass of 1.4-1.5 M$_{\odot}$ and the Roche lobe
filling secondary should have a maximum mass of 1.0 M$_{\odot}$.
The fact that many q values were pegged at the hard upper limit
implies that the primary might have an unusually low mass for a
neutron star and certainly does not support the early suggestions
that the system might contain a black hole. The disk outer radius
was also truncated to 0.5 to easily fit within the Roche lobe of the
primary. We get 0.4-0.5 for disk outer radius. The emissivity ratio
of inner/outer rim (2-3.5) roughly agrees with the values obtained
by \cite{hellmason}. Also, as expected, the fit is able to
constrain the location of zero phase very accurately. Finally, 
we find that the X-ray heating component provides $\sim$ 10\% of the
optical (U,B) flux at maximum. For error estimates on these parameter
values we refer the reader to Figure 4 and Figure 5, which show the
distributions of resulting parameters from our fits.    

It is interesting to note that if we take the H/R ratio in the
direction of the L1 point to be $\sim$ 0.3 (Figure 5) and take the
most probable value of 0.7 for the mass-ratio, we find using the
standard formula by (Eggleton, 1983) that the secondary is almost
completely in the X-ray shadow due to the disk rim (R2/a $\sim$ 0.35).
This further suggests that at the time of our observations the
X-ray heating of the secondary was not significant. As noted earlier
we get 10\% modulation due to X-ray heating from our fits, while
\cite{thorst} report an amplitude of $\sim$ 25\%. Clearly something
has changed in the way the system accretes between these two epochs.
The overall brightness has however remained roughly the same. 
Thorstensen (1987) gives a mean brightness of V=18.8. This is
consistent with the phases of maximum brightness in our light curves.
It is thus feasible that the light curves we observed could be
produced by 'extra' vertical structure imposed on a 'clean' disk
seen by \cite{thorst}. What has caused this change in the appearance
of the accretion disk is unclear. One possibility might be that there
has been a change in the amount of accreting matter leaving the
L1-point. This is supported by the RXTE All Sky Monitor lightcurve
of 4U 1957+115, which shows erratic variations on timescales of months 
between 20 and 70 mCrab. Another possibility is that the system evolves on a 
'superorbital' period as has been observed in a number of systems
(30.5 d cycle in LMC X-4 \cite{lang}, 35 d cycle in Her X-1
\cite{tananbaum}, 60 d cycle in SMC X-1 \cite{gruber}, 77.7 d cycle
in Cyg X-2 \cite{wijnands} and 106 d cycle in M33-X8 \cite{dubus}).

Recently, \cite{hakala98} reported their results of a monitoring 
campaign of MS1603+26, a possible short period LMXB. They found out that
this system also shows remarkably different optical light curves
at different epochs. Furthermore, \cite{callagrin} have reported their
results on 4U 1916-05, which also show evolving light curves. These
results seem to point out that evolving optical light curves are a
much more common phenomenon in LMXB's than previously thought. This
implies that accretion disks in these systems also seem to be less
static in binary frame than usually thought. On the other hand,
some systems similar to 4U 1957+115 (e.g. X1822-371) , which show
almost identical light curve shape, have remarkably
stable light curves \cite{mascor82},\cite{hellmason}. What is the reason 
behind this behaviour remains an open question, the answer to which 
could provide new means to probe the accretion flow and the nature of 
the compact object in these systems.      

\section{Conclusions}

Our UBVRI photometry of 4U1957+115 supports the 9.33 hours period found by \cite{thorst}.
We observe, however, significant changes in the light curve parameters when compared to
those measured by \cite{thorst}. First of all, the amplitude of modulation is $\sim$ 0.5-0.6
mag compared to $\sim$ 0.23 mag observed by \cite{thorst}. Secondly, we find that the 
light curve is no longer sinusoidal, but displays clear asymmetry. We also find, that
the source is at its bluest in the phase of {\it minimum
brightness}. Our model fits yield some estimates for the system
parameters and suggest significant geometrical thickness for the
non-axisymmetric disk. Further observations are strongly encouraged to monitor the
changes in the light curve of 4U 1957+115 and other LMXB's, which are
likely to show similar behaviour.

\section{Acknowledgements}

These observations were obtained at the Nordic Optical Telescope
(NOT), Observatorio de Roque de los Muchachos, La Palma. We would
like to thank the NOT staff for their help. We would also like to
thank Phil Charles and Keith Mason for friendly advice. PJH announces
the support from the Academy of Finland and wishes to thank MSSL for
hospitality during his extended visit. PM wishes to thank the
Finnish Cultural Foundation for financial support. GD would like to
thank PPARC for travel support to NOT and the Oxford astrophysics 
department for hospitality.

\end{document}